\documentstyle[preprint,aps]{revtex}
\begin{document}
\draft
\preprint{ hep-th/9505036 }
\title{Phase transition by curvature in three dimensional $O(N)$
sigma model}
\author{Dae-Yup Song\footnote{Electronic address:
        dsong@sunchon.sunchon.ac.kr}}
\address{Department of Physics,\\ Sunchon National University, Sunchon
540-742, Korea}
\date{ April, 1995}

\maketitle

\begin{abstract}
Using the effective potential, the large-$N$ nonlinear $O(N)$ sigma
model with the curvature coupled term is studied on $S^2\times R^1$.
We show that, for the conformally coupled case, the dynamical mass
generation of the model in the strong-coupled regime on $R^3$ takes
place for any finite scalar curvature (or radius of the $S^2$).
If the coupling constant is larger than that of the conformally
coupled case, there exist a critical curvature (radius) above (below)
which the dynamical mass generation does not take place even in
the strong-coupled regime. Below the critical curvature, the mass
generation occurs as in the model on $R^3$.
\end{abstract}


\section{Introduction}

There has been a lot of interest in quantum field theories on curved
spacetime background\cite{BiDa}. In general, quantum field theories
may be sensitive to both the local and global structure of spacetime.
The structure can alter the physical parameters in quantum loop
corrections, which can not be absorbed by simple redefinition of the
parameter after removing the ultraviolet divergences. A well-known
example is the finite temperature field theory, where the
renormalized parameters can become temperature dependent. This
phenomenon could be of particular interest for the models in which the
physical parameters get non-vanishing values through quantum loop
corrections\cite{Coleman}.

In this paper we will study the three dimensional nonlinear $O(N)$
sigma model\cite{Aref'eva} on $S^2\times R^1$ Euclidean spacetime by
evaluating the effective potential in the large-$N$
limit\cite{Coleman,Song}. In the model on $R^3$, there exist a
critical coupling constant $g_c$ which separates the strongly coupled
case (strong-coupled regime) and weakly coupled case (weak-coupled
regime), while dynamical mass generation takes place only in the
strongly coupled case \cite{Aref'eva,Song}. Due to $S^2$, the
spacetime should have nonvanishing scalar curvature\footnote{ The
author thanks Professor P. Gilkey and Dr. J.H. Park for discussions on
this point.}. For the sake of explicit evaluations, we will restrict
our discussions on the case of the canonical metric for $S^2$ and the
flat metric on $R^1$ which gives the constant scalar curvature,
$R=2/\rho^2$, with the radius $\rho$ of $S^2$. In the presence of
scalar curvature, we may add a gravitational interaction term
$\xi R n^i n_i$ $(=\xi R n^2)$ to Lagrangian, where $\xi$ is a
gravitational coupling constant and $n^i$ $(i=1,2,\cdots,N)$ is a
boson field. Taking the above arguments into consideration, the
Lagrangian of the nonlinear $O(N)$ sigma model on $S^2\times R^1$ may
be written as:
\begin{eqnarray}
L=\int d\theta d\varphi dz\rho^2\sin\theta
  [&&\frac{1}{\rho^2} \partial_\theta n^i\partial_\theta n^i
   +\frac{1}{\rho^2\sin^2\theta}
   \partial_\varphi n^i\partial_\varphi n_i \nonumber\\
   &&+\partial_z n^i\partial_z n_i +\xi Rn^2
     +\sigma(n^2-\frac{N}{g_0^2})],
\end{eqnarray}
where $\sigma$ is a Lagrange multiplier to coerce the constraint
$n^2=N/g_0^2$ and $g_0$ is the bare coupling constant of the model.

In the literatures\cite{BiDa} special emphases have been made on the
conformally coupled cases: $\xi=\frac{d-2}{4(d-1)}$ on $d$-dimensional
spacetime or $\xi=\frac{1}{8}$ on three dimension. On $S^d$ or
$S^{d-n}\times R^n$ there have been some analyses for other
models\cite{DruSh,RT&KM}, and the $\zeta$-function method has been
mostly used in isolating divergences of quantum loop
correction\cite{BiDa,DruSh,RT&KM,DS,O'Connor&Hu&Shen}, while we will
use the cut-off method.

In the next section we will evaluate the effective potential, $V$, on
$S^2\times R^1$ in the leading order of $N$.  For $\xi\geq
\frac{1}{8}$, the renormalized effective potential will be explicitly
found. In Sec.III, we will study whether the dynamical mass generation
takes place by investigating the stationary point in $V$. It will be
shown that for $\xi=\frac{1}{8}$ dynamical mass generation will take
place for any value of $\rho$ if $g$ is larger than $g_c$, as in the
model on $R^3$. For $\xi>\frac{1}{8}$, there exists a critical
curvature $R_c$ $(=2/\rho_c^2)$; For $R\geq R_c$, dynamical mass
generation does not occur even in the strongly coupled case. For
$\xi>\frac{1}{8}$, we found the analytic expression of $R_c$, which
shows that $R_c$ decreases monotonically as $\xi$ increases. The final
section will be devoted on discussions.

\section{Effective potential}

In the leading order of the $1/N$ expansion, the effective potential
for constant $\sigma$ is given by the tree and one-loop diagrams with
external $\sigma$ lines\cite{Coleman&Weinberg,Coleman}. Or, in order
to make use of functional integral\cite{Jackiw}, one can write the
Lagrangain density with the quantum mechanical angular momentum
operator ${\bf L}$ as follows;
\begin{equation}
{\cal L}=n^i Dn_i-N\sigma/g_0^2,
\end{equation}
where
\begin{equation}
D=-\partial_z^2 +\frac{{\bf L}^2}{\rho^2} +\frac{2\xi}{\rho^2}
 +\sigma.
\end{equation}
By performing the Gaussian functional integral, one can find the
effective potential per unit volume as
\begin{equation}
\frac{V_0}{N}=-\frac{\sigma}{g_0^2}
+\frac{1}{4\pi \rho^2}(\text{Tr}\ln D+C).
\end{equation}
$C$ is a constant which may arise from the functional integral and it
will be fixed by demanding that $V_0\mid_{\sigma=0}=0$. As is
well-known, the diagrams which have $\sigma$ propagators as internal
lines give contributions of next to the leading order in the $1/N$
expansion. Therefore, the effective potential of the leading order per
unit volume can be written as;
\begin{equation}
\frac{V}{N}=-\frac{\sigma}{g_0}+\frac{1}{4\pi\rho^2}\sum_{l=0}^{I}
   \int_{|k|<\Lambda_1} \frac{dk}{2\pi}(2l+1)
   \ln(1+\frac{\sigma}{k^2+\frac{l(l+1)}{\rho^2}+\frac{2\xi}{\rho^2}
   +i\epsilon}).
\end{equation}
In Eq.(5), we introduce the cut-off $\Lambda_1$ and $I$. While the
mass dimension of $\Lambda_1$ is that of a momentum as usual, the $I$
is a pure number since it is a cut-off for the quantum number of the
operator ${\bf L}$. If sapcetime is a product of two manifolds with
different topologies, it is necessary to introduce different cut-offs.
Although this should happen in finite-temperature field theories, this
could be implicit if summation or integral along some directions are
finite.

In order to compare the effective potential in Eq.(5)  with that on
$R^3$, one can use the following formula:
\begin{equation}
\sum_{l=0}^N f(l)=\frac{1}{2}f(0)+\int_0^{N+1}f(x)dx
                  +\sum_{l=0}^N\int_0^1 f'(x+l)(x-\frac{1}{2})dx
                  -\frac{1}{2}f(N+1),
\end{equation}
which comes from the Euler-Maclaurin formula
\begin{equation}
\int_0^1 f(x+l) dx =\frac{1}{2}(f(l+1)+f(l))-\int_0^1 f'(x+l)
                  (x-\frac{1}{2})dx
\end{equation}
for any differentiable function $f(x)$.
Making use of the equality in Eq.(6), the effective potential can be
written as
\begin{eqnarray}
\frac{V}{N}&=&-\frac{\sigma}{g_0^2}
   +\frac{1}{8\pi^2\rho^2}\int_{|k|<\Lambda_1}
           \int_{x=0}^{I} dx (2x+1)
     \ln(1+\frac{\rho^2\sigma}{(x+\frac{1}{2})^2+2(\xi-\frac{1}{8})
        +\rho^2k^2+i\epsilon})\nonumber\\
 &&+\frac{1}{16\pi^2 \rho^2} \int_{-\infty}^{\infty} dk
   \ln(1+\frac{\rho^2\sigma}{\rho^2 k^2 +2\xi +i\epsilon})\nonumber\\
 &&+\frac{1}{8\pi^2}\int_{-\infty}^{\infty}dk\sum_{l=0}^\infty
        \int_0^1 dx (x-\frac{1}{2})
   \left[
    \begin{array}{l}
    \frac{2}{R^2}\ln (1+\frac{\rho^2\sigma}{(x+\frac{1}{2}+l)^2
             +2(\xi-\frac{1}{8})+\rho^2k^2+i\epsilon})\\
    -\frac{4\sigma(x+\frac{1}{2}+l)^2}
      {[(x+\frac{1}{2}+l)^2+2(\xi-\frac{1}{8})+\rho^2k^2+i\epsilon]
           [(x+\frac{1}{2}+l)^2+2(\xi-\frac{1}{8})+\rho^2k^2
                       +\rho^2\sigma+i\epsilon] }
    \end{array}\right]\nonumber\\
 &&+O(1/\Lambda_1)+O(1/I)
\end{eqnarray}
It is convenient to divide $V$ into two pieces so that
\begin{equation}
V=V_\xi+V_\Delta+O(1/\Lambda_1)+O(1/I),
\end{equation}
where
\begin{equation}
\frac{V_\xi}{N}=-\frac{\sigma}{g_0^2}
   +\frac{1}{8\pi^3}\int_{|k|<\Lambda_1}dk
   \int_{k_2=0}^{\Lambda_2} dk_2~2\pi k_2
   \ln(1+\frac{\sigma}{k^2+k_2^2+\frac{2(\xi-\frac{1}{8})}{\rho^2}
        +i\epsilon})
\end{equation}
and
\begin{eqnarray}
\frac{V_\Delta}{N}&=&
   \frac{1}{16\pi^2 \rho^2} \int_{-\infty}^{\infty} dk
      \ln(1+\frac{\sigma}{k^2 +\frac{2\xi}{\rho^2}+i\epsilon})
                 \nonumber\\
 &&-\frac{1}{4\pi^2\rho^2}\int_0^\frac{1}{2} dt~t
   \int_{-\infty}^\infty dk
     \ln(1+\frac{\sigma}{k^2+(\frac{t}{\rho})^2
      +\frac{2}{\rho^2}(\xi-\frac{1}{8})+i\epsilon})\nonumber\\
 &&+\frac{1}{8\pi^2}\sum_{l=0}^\infty\int_{-\frac{1}{2}}^{\frac{1}{2}}
       dt~t     \int_{-\infty}^{\infty}dk
 \left[
    \begin{array}{l}
    \frac{2}{R^2}\ln (1+\frac{\sigma}{k^2+\frac{(t+l)^2}{\rho^2}
          +\frac{2(\xi-\frac{1}{8})}{\rho^2}+i\epsilon})\\
    -\frac{4\sigma(t+l)^2/\rho^4}
          {[k^2+\frac{(t+l)^2}{\rho^2}
                 +\frac{2(\xi-\frac{1}{8})}{\rho^2}+i\epsilon]
           [k^2+\frac{(t+l)^2}{\rho^2}+\sigma
                 +\frac{2(\xi-\frac{1}{8})}{\rho^2}+i\epsilon] }
    \end{array}\right].
\end{eqnarray}
In Eq.(10), the $\Lambda_2$ whose dimension is of momentum is
defined as
\[\Lambda_2=\frac{I+\frac{1}{2}}{\rho}.\]
{}From now on we will restrict our attention on the case
$\xi\geq\frac{1}{8}$, for which logarithm functions or their integrals
in Eqs.(8,10,11) are well defined even when $\epsilon=0$. For
discussions on the case of $\xi<\frac{1}{8}$, we should develop some
analytic continuations for logarithm functions, which is beyond the
scope of this paper.

Making use of the formulae\cite{Rtable}
\begin{equation}
\int_{-\infty}^\infty \ln\frac{\alpha^2+x^2}{\beta^2+x^2}dx
   =2(|\alpha|-|\beta|)\pi,
\end{equation}
\begin{equation}
\int_{-\infty}^\infty \frac{dx}{(\gamma+x^2)(\delta+x^2)}
 =\frac{\pi}{\sqrt{\gamma\delta}(\sqrt{\gamma}+\sqrt{\delta})}
 =\frac{\pi}{\sqrt{\gamma\delta}}\frac{\sqrt{\delta}-\sqrt{\gamma}}
                                     {\delta-\gamma},
\end{equation}
we can find a simpler form of $V_\Delta$;
\begin{eqnarray}
V_\Delta
 &=&-\frac{1}{2\pi \rho^2}\int_0^{\frac{1}{2}}dt~t
   [\sqrt{(\frac{t}{\rho})^2+\frac{2(\xi-\frac{1}{8})}{\rho^2}+\sigma}
      -\sqrt{(\frac{t}{\rho})^2+\frac{2(\xi-\frac{1}{8})}{\rho^2}}]
      \nonumber\\
 &&+\frac{1}{8\pi \rho^2}[\sqrt{\frac{2\xi}{\rho^2}+\sigma}
          -\frac{\sqrt{2\xi}}{\rho}]
      \nonumber\\
 &&+\frac{1}{2\pi}\sum_{l=1}^\infty\int_{-\frac{1}{2}}^{\frac{1}{2}}
 dt~ t\left[
 \begin{array}{l}
  \frac{1}{\rho^2}[\sqrt{\frac{(t+l)^2}{\rho^2}
    +\frac{2(\xi-\frac{1}{8})}{\rho^2}+\sigma}
      -\sqrt{\frac{(t+l)^2}{\rho^2}
              +\frac{2(\xi-\frac{1}{8})}{\rho^2}} ]      \\
   -\frac{\sigma(t+l)^2}
      {\rho\sqrt{[(t+l)^2+2(\xi-\frac{1}{8})+\rho^2\sigma]
             [(t+l)^2+2(\xi-\frac{1}{8})]}}\\
   \times\frac{1}{\sqrt{[(t+l)^2+2(\xi-\frac{1}{8})+\rho^2\sigma}
         +\sqrt{[(t+l)^2+2(\xi-\frac{1}{8})}}
 \end{array}\right] \\
 &=&-\frac{1}{2\pi \rho^3}\int_0^{\frac{1}{2}}dt~t
   [\sqrt{t^2+2(\xi-\frac{1}{8})+\rho^2\sigma}
      -\sqrt{t^2+2(\xi-\frac{1}{8})}]
      \nonumber\\
 &&+\frac{1}{8\pi \rho^3}[\sqrt{2\xi+\rho^2\sigma}-\sqrt{2\xi}]
      \nonumber\\
 &&+\frac{1}{2\pi \rho^3}\sum_{l=1}^\infty
  \int_{-\frac{1}{2}}^{\frac{1}{2}}dt~ t\left[
  \begin{array}{l}
   [\sqrt{(t+l)^2+2(\xi-\frac{1}{8})+\rho^2\sigma}
       -\sqrt{(t+l)^2+2(\xi-\frac{1}{8})}]\\
   \times[1-
    \frac{(t+l)^2}{\sqrt{(t+l)^2+2(\xi-\frac{1}{8})+\rho^2\sigma}
    \sqrt{(t+l)^2+2(\xi-\frac{1}{8})}}]
  \end{array}\right],
\end{eqnarray}
which clearly shows that $V_\Delta$ is finite for any $\rho$,
$\sigma$. Furthermore $V_\Delta$ is zero in the limit $\rho$
approaches to infinity (in the $R^3$ limit), which shows that it is an
effect of topology. In this limit, the integral of Eq.(10) for $V_\xi$
which diverges with infinite cut-offs, in fact, corresponds to that of
the model on $R^3$:
%\begin{equation}
\[
\frac{1}{8\pi^3}
   {\int dk \int dk_2}_{k_2>0,k^2+k_2^2 < \Lambda^2}~2\pi k_2
   \ln(1+\frac{\sigma}{k^2+k_2^2+i\epsilon}).
\]
%\end{equation}
The difference in the $R^3$-limit is that we have two cut-offs,
because our spacetime $S^2\times R^1$ is the product of two spaces
with different topologies. As mentioned earlier this should happen,
for example, in finite-temperature field theory, if we use cut-off
regularization. In renormalizing the model on $S^2\times R^1$ we will
assume - as usual - that we could deform the momentum space of
integration.
That is, the $V_\xi$ which does not vanish in the $R^3$-limit could be
written as:
\begin{eqnarray}
\frac{V_\xi}{N}&=&-\frac{\sigma}{g_0^2}
+\frac{1}{8\pi^3}
   {\int dk \int dk_2}_{k_2>0,k^2+k_2^2 < \Lambda^2}~2\pi k_2
   \ln(1+\frac{\sigma}{k^2+k_2^2+\frac{2(\xi-\frac{1}{8})}{\rho^2}
        +i\epsilon}) \\
&=&-\frac{\sigma}{g_0^2} +\frac{\sigma\Lambda}{2\pi^2}-\frac{1}{6\pi}
     \sqrt{[\frac{2(\xi-\frac{1}{8})}{\rho^2}+\sigma]^3}
     +\frac{1}{3\pi \rho^3}\sqrt{2(\xi-\frac{1}{8})^3}\nonumber\\
 &&+O(1/\Lambda),
\end{eqnarray}
which shows that the divergence of the potential on $S^2\times R^1$
(or $\Lambda$ dependence) is the same as that on $R^3$ and we could
renormalize the model on $S^2\times R^1$ by using the relation on
$R^3$\cite{Song}:
\begin{eqnarray}
\frac{1}{g_0^2}&=&\frac{1}{g^2}+\int_{|p|<\Lambda}
\frac{d^3p}{(2\pi)^3}\frac{1}{p^2+M^2}\\
&=&\frac{1}{g^2}+ \frac{\Lambda}{2\pi}-\frac{M}{4\pi}+O(1/\Lambda),
\end{eqnarray}
where $M$ is the renormalization mass.

The fact that topological change of spacetime does not give rise to
new counterterms in the renormalization has been found explicitly for
some cases (for example, see Ref. \cite{DruSh}) and a more general
discussion has been given in Ref.\cite{Banach}.

Now we have the renormalized effective potential for $\xi\geq
\frac{1}{8}$
\begin{equation}
\frac{V}{N}=-\frac{\sigma}{g^2}+\frac{\sigma M}{4\pi}-\frac{1}{6\pi}[
   \sqrt{[\frac{2(\xi-\frac{1}{8})}{\rho^2}+\sigma]^3}
     +\frac{1}{3\pi \rho^3}\sqrt{2(\xi-\frac{1}{8})^3}]
 +\frac{V_\Delta}{N}+O(1/\Lambda),
\end{equation}
which, in the $R^3$-limit, reduces to the potential on $R^3$, $V_0$
\cite{Song}:
\begin{equation}
\frac{V_0}{N}=-\frac{\sigma}{g^2}+\frac{\sigma M}{4\pi}
         -\frac{\sigma^{3/2}}{6\pi} +O(1/\Lambda).
\end{equation}
As a preliminary of the next section, it may be good to recapitulate
the dynamical properties of the model on $R^3$. The first derivative
of $V_0$ with respect to $\sigma$
\[\frac{1}{N} \frac{\partial V_0}{\partial \sigma}=
  \frac{M}{4\pi}-\frac{1}{g^2}-\frac{\sqrt{\sigma}}{4\pi}+O(1/\Lambda)
\]
shows that there exist a global stationary point only when $g^2 >
\frac{4\pi}{M}(=g_c^2)$. That is, dynamical mass generation takes
place only in the strongly coupled case ($g^2 > g_c^2$), and $g_c$
is the critical coupling constant. The dynamically generated mass is
$\sqrt{\sigma_0}$ $(=M-\frac{4\pi}{g^2})$ which reduces to zero as
$g$ goes to $g_c$ \cite{GRV}.

\section{Phase Structure}

Though the forms of $V_0$ and $V$ differ by $V_\Delta$ which is rather
complicated, they share the facts that $V_0\mid_{\sigma=0}=
V\mid_{\sigma=0}=0$ and
\[ V\simeq V_0\simeq  -\frac{\sigma^{3/2}}{6\pi}
        ~~~\text{for large } \sigma;\]
That is, they start from 0 and decrease to $-\infty$ as $\sigma$
increases to $\infty$. To see the shape of $V$, we evaluate the first
derivative of $V$ with respect to $\sigma$;
\begin{equation}
\frac{\rho}{N}\frac{\partial V}{\partial \sigma}=
(\frac{M}{4\pi}-\frac{1}{g^2})\rho - f(\xi;y),
\end{equation}
where
\begin{eqnarray}
f(\xi;y)
      &=&\frac{1}{4\pi}\sqrt{2(\xi-\frac{1}{8})+y}
         -\frac{1}{16\pi}\frac{1}{\sqrt{2\xi+y}}
         +\frac{1}{4\pi}\int_0^{\frac{1}{2}}dt~t
           \frac{1}{\sqrt{t^2+2(\xi-\frac{1}{8})+y}}\nonumber\\
 &&-\frac{1}{4\pi}\sum_{l=1}^\infty
 \int_{-\frac{1}{2}}^{\frac{1}{2}}dt~t
 \left[
   \begin{array}{l}
    \frac{1}{\sqrt{(t+l)^2+2(\xi-\frac{1}{8})+y}}
      [1-\frac{(t+l)^2}{\sqrt{(t+l)^2+2(\xi-\frac{1}{8})+y}
                        \sqrt{(t+l)^2+2(\xi-\frac{1}{8})} }]    \\
   +\frac{(t+l)^2}{\sqrt{[(t+l)^2+2(\xi-\frac{1}{8})+y]^3
                          [(t+l)^2+2(\xi-\frac{1}{8})]}  }       \\
      \times[\sqrt{(t+l)^2+2(\xi-\frac{1}{8})+y}-\sqrt{
                                   (t+l)^2+2(\xi-\frac{1}{8})}]
   \end{array} \right]  \nonumber\\
      &=&\frac{1}{4\pi}\sqrt{2(\xi-\frac{1}{8})+y}
         -\frac{1}{16\pi}\frac{1}{\sqrt{2\xi+y}}
         +\frac{1}{4\pi}\int_0^{\frac{1}{2}}dt~t
           \frac{1}{\sqrt{t^2+2(\xi-\frac{1}{8})+y}}\nonumber\\
 &&-\frac{2(\xi-\frac{1}{8})+y}{4\pi}\sum_{l=1}^\infty
 \int_{-\frac{1}{2}}^{\frac{1}{2}}dt~t
   \frac{t}{[(t+l)^2+2(\xi-\frac{1}{8})+y]^{3/2}}
\end{eqnarray}
with $y=\rho^2\sigma$.
Making use of the formulae in the appendix, a further simplification
of $f$ is possible;
\begin{eqnarray}
f(\xi;y)=\frac{1}{4\pi}{\cal F}_{\frac{1}{2}}
           (\sqrt{2(\xi-\frac{1}{8})+y})
&=&\frac{1}{4\pi}[{\cal F}(2\sqrt{2(\xi-\frac{1}{8})+y})
             -{\cal F}(\sqrt{2(\xi-\frac{1}{8})+y})],
\end{eqnarray}
where for $\alpha >0$ ${\cal F}_{\frac{1}{2}}(\alpha)$ and
${\cal F}(\alpha)$ are defined by
\begin{eqnarray}
{\cal F}_{\frac{1}{2}}(\alpha) &=&
         \sum_{l=0}^\infty[1-\frac{l+\frac{1}{2}}
               {\sqrt{\alpha^2+(l+\frac{1}{2})^2}}],\nonumber\\
{\cal F}(\alpha)&=&\sum_{l=1}^\infty[1
           -\frac{1}{\sqrt{1+(\frac{\alpha}{l})^2}}].
\end{eqnarray}
It is easy to find that ${\cal F}(0)=0$ and
${\cal F}_{\frac{1}{2}}(\alpha)$ monotonically increases as $\alpha$
increases, which proves that if there is a stationary point of $V$
then it must be global. Since the dynamical mass generation is denoted
by the presence of stationary point of the effective potential, the
above facts reveal two properties of the model on $S^2\times R^1$
for $\xi\geq \frac{1}{8}$. The first is that if dynamical mass
generation takes place it occurs only in the strong-coupled regime
($g^2 >g_c^2$), since $f(\xi;y)$ is never less than 0. Secondly, even
in the strong-coupled regime dynamical mass generation takes place
only if $\rho$ $(R)$ of 2-sphere is larger (smaller) than $\rho_c$
($R_c$).
Since $f(\xi;y)$ is monotonically increasing function of $y$ for a
fixed $\xi$, the critical condition is given by
\begin{equation}
\frac{\partial V}{\partial \sigma}\mid_{\sigma=0}=0,
\end{equation}
which gives the $\rho_c$ as;
\begin{equation}
\rho_c \sqrt{\sigma_0}={\cal F}_{\frac{1}{2}}(\sqrt{2\xi-\frac{1}{4}})
               ={\cal F}(2\sqrt{2\xi-\frac{1}{4}})
                  - {\cal F}(\sqrt{2\xi-\frac{1}{4}}).
\end{equation}

The smallest $\rho_c$ (largest $R_c$) which is in the the conformal
coupling case $(\xi=\frac{1}{8})$, is $0$ ($\infty$) since
${\cal F}(0)$ is zero. In other words, the dynamical mass generation
takes place for any size of sphere in the strong-coupled regime of
the conformal coupling case.

\section{Conclusion}

We have studied the large-$N$ nonlinear $O(N)$ sigma model on
$S^2\times R^1$ by evaluating the effective potential with cut-off
method. By analysing the (renormalized) effective potential for
$\xi\geq \frac{1}{8}$, we find that in the strongly coupled case there
exists critical size or curvature of 2-sphere. Even in the strongly
coupled case the dynamical mass generation does not take place when
the radius (curvature) of two sphere is smaller (larger) than the
critical one.

As the change of temperature may cause phase transition in
finite-temperature field theory, it has been known that the change of
curvature of background spacetime could give rise to phase transition
\cite{Ford&Toms,O'Connor&Hu&Shen}, while our results provide another
explicit example. The smallest critical radius (largest critical
curvature) for the model we considered is $0$ ($\infty$) in conformal
(gravitational) coupling case. Thus on $S^2\times R^1$ dynamical mass
generation of the model on $R^3$ always takes place in the conformal
coupling case.

The value of $\sigma$ at which the potential is stationary is the
square of the dynamically generated mass. Though we cannot
analytically find the value of dynamically generated mass, the fact
that $f(\xi;y)$ is a monotonically increasing function of $y$ suggests
that the phase transition would be of second order, since the mass
continuously approach zero as $\rho$ goes to $\rho_c$ from the above
($\rho>\rho_c$). For the conformal coupling case $(\xi=1/8)$,
$f(\frac{1}{8};y)$ is also a monotonically increasing function of
$y$. Therefore, at $\xi=1/8$ the dynamically generated mass
approaches 0 as $g$ goes to $g_c$, which is support the
ansatz made in Ref. \cite{GRV}.

It would be interesting to analyse the case $\xi<\frac{1}{8}$ which is
beyond the scope of this paper. It would be also of great interest to
analyse the model on different topologies. A similar analysis of the
model on $S^2$ will be published elsewhere\cite{Song2}.

\acknowledgments

The author thanks Professor L.H. Ford for fruitful discussion on his
works and Professor G. 't Hooft for discussion on treatment of
cut-offs. He also thanks Dr. W. Bietenholz for kind comments on the
first version of this paper. This work was supported in part by the
KOSEF and also by the U.S. Department of Energy under Contract No.
DE-AC02-76-ER03130.

\section*{Appendix}

In Sec. III the following facts are used;
\begin{equation}
\sum_{l=1}^\infty\int_{-\frac{1}{2}}^{\frac{1}{2}} dt
 \frac{t}{[(t+l)^2+\alpha^2]^{3/2}}= I_1+ I_2,
\end{equation}
where
\begin{equation}
I_1=\sum_{l=1}^\infty\int_{l-\frac{1}{2}}^{l+\frac{1}{2}}
  \frac{y}{[y^2+\alpha^2]^{3/2}} dy
     =\frac{1}{\sqrt{\frac{1}{4}+\alpha^2}}
\end{equation}
and
\begin{eqnarray}
\alpha^2 I_2
&=&\sum_{l=1}^\infty l[\frac{l-\frac{1}{2}}
            {\sqrt{\alpha^2+(l-\frac{1}{2})^2}}
       -\frac{l+\frac{1}{2}}{\sqrt{\alpha^2+(l+\frac{1}{2})^2}}]
    \nonumber\\
&=&\frac{1}{2\sqrt{\alpha^2+\frac{1}{4}}} + \lim_{N\rightarrow\infty}
   [\sum_{l=1}^N
   \frac{l+\frac{1}{2}}{\sqrt{\alpha^2+(l+\frac{1}{2})^2 }}
   -(N+1)\frac{N+\frac{3}{2}}{\sqrt{\alpha^2+(N+\frac{3}{2})^2}}]
    \nonumber\\
&=&\sum_{l=0}^\infty
   [\frac{l+\frac{1}{2}}{\sqrt{\alpha^2+(l+\frac{1}{2})^2}} -1]
=-{\cal F}_{\frac{1}{2}}(\alpha).
\end{eqnarray}
${\cal F}_{\frac{1}{2}}(\alpha)$ and ${\cal F}(\alpha)$ are related
as follows;
\begin{equation}
{\cal F}_{\frac{1}{2}}(\alpha) ={\cal F}(2\alpha)-{\cal F}(\alpha).
\end{equation}


\begin{references}
\bibitem{BiDa} N.D. Birrell and P.C.W Davies, {\em Quantum Fields
    in Curved Space} (Cambridge Univ. Press, Cambridge, 1982).
\bibitem{Coleman} S. Coleman, {\em Aspects of Symmetry}
     (Cambridge Univ. Press, Cambridge, 1985).
\bibitem{Aref'eva} I. Ya. Aref'va, Teo. Mat. Fiz. {\bf 29}, 147
    (1976); {\bf 36}, 159 (1978); Ann. Phys. (N.Y.) {\bf 117},
    393 (1979).
\bibitem{Song} D.Y. Song, Pys. Rev. D {\bf 49}, 6794 (1994).
\bibitem{DruSh} I.T. Drummond, Nucl. Phys. B {\bf 94}, 115 (1975);
    I.T. Drummond and G.M. Shore, Phys. Rev. D {\bf 19}, 1134 (1979);
    G.M. Shore, Phys. Rev. D {\bf 21}, 2226 (1980).
\bibitem{RT&KM} I.H. Russell and D.J. Toms, Class. Quantum Grav.
    {\bf 4}, 1357 (1987);
    R. Kantowski and K.A. Milton, Phys. Rev. D {\bf 35}, 549 (1987).
\bibitem{DS} B.S. DeWitt, Phys. Rep. {\bf 19}, 295 (1975);
    S. W. Hawking, Comm. Math. Phys. {\bf 55}, 133 (1977);
    J.S. Dowker, Comm. Math. Phys. {\bf 162}, 633 (1994);
    G. Cognola, K. Kirsten and S. Zerbini, Phys. Rev. D {\bf 48}, 790
    (1993), and references therein.
\bibitem{O'Connor&Hu&Shen} D.J. O'Connor, B. L. Hu, and T.C. Shen,
   Phys. Lett. {\bf 130B}, 31; T.C. Shen, B.L. Hu, and D.J. O'Connor,
   Phys. Rev. D {\bf 31}, 2401 (1985).
\bibitem{Coleman&Weinberg} S. Coleman and E. Weinberg, Phys. Rev. D
   {\bf 7}, 1888 (1973).
\bibitem{Jackiw} R. Jackiw, Phys. Rev. D {\bf 9}, 1686 (1974).
\bibitem{Rtable} For example, see I.S. Gradshteyn and I.M. Ryzhik,
   {\em Table of
   of integrals, Series, and Products} (Academic Press, Boston, 1994).
\bibitem{Banach} R. Banach, J. Phys. A: Math. Gen. {\bf 13}, L365
   (1980).
\bibitem{GRV} S. Guruswamy, S.J. Rajeev and P. Vitale, Nucl. Phys. B
   {\bf 438}, 491 (1995).
\bibitem{Ford&Toms} L.H. Ford and D.J. Toms, Phys. Rev. D {\bf 25},
   1510 (1982).
\bibitem{Song2} D.Y. Song, in preparation.
\end{references}
\end{document}